\documentclass[superscriptaddress,reprint]{revtex4-1}
\usepackage{xcolor}
\usepackage{graphicx}
\usepackage{dcolumn}
\usepackage{tikz}
\usepackage{url}
\usepackage{mathptmx}
\usepackage{amsmath,amsfonts}
\usepackage{algorithmic}
\usepackage{algorithm}
\usepackage{array}
\usepackage{textcomp}
\usepackage{verbatim}
\usepackage[hidelinks]{hyperref}

\usetikzlibrary{shapes.geometric, arrows, positioning}

\tikzstyle{startstop} = [rectangle, rounded corners, 
minimum width=4cm, 
minimum height=1cm,
text centered, 
draw=black, 
fill=red!30]

\tikzstyle{io} = [trapezium, 
trapezium stretches=true, 
trapezium left angle=70, 
trapezium right angle=110, 
minimum width=3cm, 
minimum height=1cm, text centered, 
draw=black, fill=blue!30]

\tikzstyle{process} = [rectangle, 
minimum width=4cm, 
minimum height=1cm, 
text centered, 
text width=4cm, 
draw=black, 
fill=orange!30]

\tikzstyle{decision} = [diamond,
aspect =2,
minimum width=3cm, 
minimum height=1cm, 
text centered, 
text width = 3cm,
draw=black, 
fill=green!30]
\tikzstyle{arrow} = [thick,->,>=stealth]

\begin{document}

\title{Using Old Laboratory Equipment with Modern Web-of-Things Standards: \\a  Smart Laboratory with LabThings Retro}

\author{Samuel McDermott}
\email[Email: ]{sjm263@cam.ac.uk}
\affiliation{Cavendish Laboratory, University of Cambridge, UK}
\author{Jurij Kotar}
\affiliation{Cavendish Laboratory, University of Cambridge, UK}
\author{Joel Collins}
\affiliation{Department of Physics, University of Bath, UK}
\author{Leonardo Mancini}
\affiliation{Cavendish Laboratory, University of Cambridge, UK}
\author{Richard Bowman}
\affiliation{Department of Physics, University of Bath, UK}
\author{Pietro Cicuta}
\affiliation{Cavendish Laboratory, University of Cambridge, UK}

\begin{abstract}
There has been an increasing, and welcome, Open Hardware trend towards science teams building and sharing their designs for new instruments. These devices, often built upon low-cost microprocessors and micro-controllers, can be readily connected to enable complex, automated, and smart experiments. When designed to use open communication web standards, devices from different laboratories and manufacturers can be controlled using a single protocol, and even communicate with each other.  However, science labs still have a majority of old, perfectly functional, equipment which tends to use older, and sometimes proprietary,  standards for communications.  In order to encourage the continued and integrated use of this equipment in modern automated experiments, we develop and demonstrate LabThings Retro.  This allows us to retrofit old instruments to use modern web-of-things standards, which we demonstrate with closed-loop feedback involving an optical microscope, digital imaging and fluid pumping.    
\end{abstract}

\date{\today}

\maketitle

\section{Introduction}
Future discoveries in a wide context of  biological experiments will depend on increasingly ``smart'' laboratories. Much biological research relies on  experiments aiming to understand casual relationships in systems with very large parameter spaces to explore, and sources of complication that  require high throughput and robust data. Exploring  large parameter spaces whilst maintaining reproducibility often becomes a challenge for the human operator. As the tasks of such an experiment are repeated over long periods of time, they can be tedious, prone to human error, and not straightforward to standardize and replicate across labs.   Currently, human operators are often  necessary in experimental pipelines to identify events of interest, to then trigger corresponding actions on equipment. This tends to discourage explorations of systems that don't fit well with human timescales, e.g. it is difficult for humans to intervene in processes that are either too fast or too slow, or where the trigger signal needs to be filtered from noise or from multiple inputs. The only feasible way to obtain data for these complex and ``unfriendly'' systems is through experimental pipelines that exploit automation and allow feedback loops. 

Many new laboratory devices and software, in particular open-source hardware approaches, are reflecting this paradigm shift. Equipment such as the liquid handling robot Opentrons~\cite{noauthor_opentrons_nodate} and measurement devices such as OpenFlexure~\cite{McDermott2022,collins_robotic_2020}  and UC2~\cite{diederich_versatile_2020} can now perform reproducible automation protocols~\cite{ouyang_opensource_2022}. Large datasets can now be analysed much more efficiently using software such as Napari~\cite{ahlers_napari_2023}, CellProfiler~\cite{stirling_cellprofiler_2021}, and ImJoy~\cite{ouyang_imjoy_2019}. By combining these technologies, it is possible to develop custom and affordable `smart' laboratories, where feedback loops can change the experiment based on previously acquired data. 

However, many laboratories have old equipment. Although mechanically and electrically working, these do not have the modern communication interfaces required for any sort of smart experiment. Some contemporary equipment also does not come with a communication interface. We can think of three categories for these devices:
\begin{enumerate}
    \item \textbf{No external communications at all.} For example, units that might just be turned on or off with a switch, e.g. hotplates, rocking tables, lamps, ultrasonic baths. 
    \item \textbf{Controlled externally,  using open (documented) communication protocols}, typically  serial protocols and connectors such as  USB or RS-232.
    \item \textbf{Controlled externally, using  proprietary and closed protocols.}  For example, they require closed-source drivers and un-documented software.
\end{enumerate}

We show how devices which are in Category 1 and 2, and depending on hardware constraints some in Category 3, can be retrofitted so that they can be integrated into smart experiments. Our demonstration uses low-cost, open source, hardware and software.   

In the case of Category 3,  tools like pyautogui~\cite{sweigart_pyautogui_2023} have been shown as useful, if crude, workarounds by simulating interactions with the proprietary Graphical User Interfaces~\cite{bertaux_enhancing_2022}. Otherwise one has to `reverse engineer' the communication protocol, which is not something we describe here. 

Unfortunately in some cases, the limitations of proprietary systems are truly encrypted and/or linked to proprietary hardware acquisition cards, and this means that it is not in general straightforward to integrate such devices into smart laboratory ecosystems. 

\begin{figure}
    \centering
    \includegraphics[width=\columnwidth]{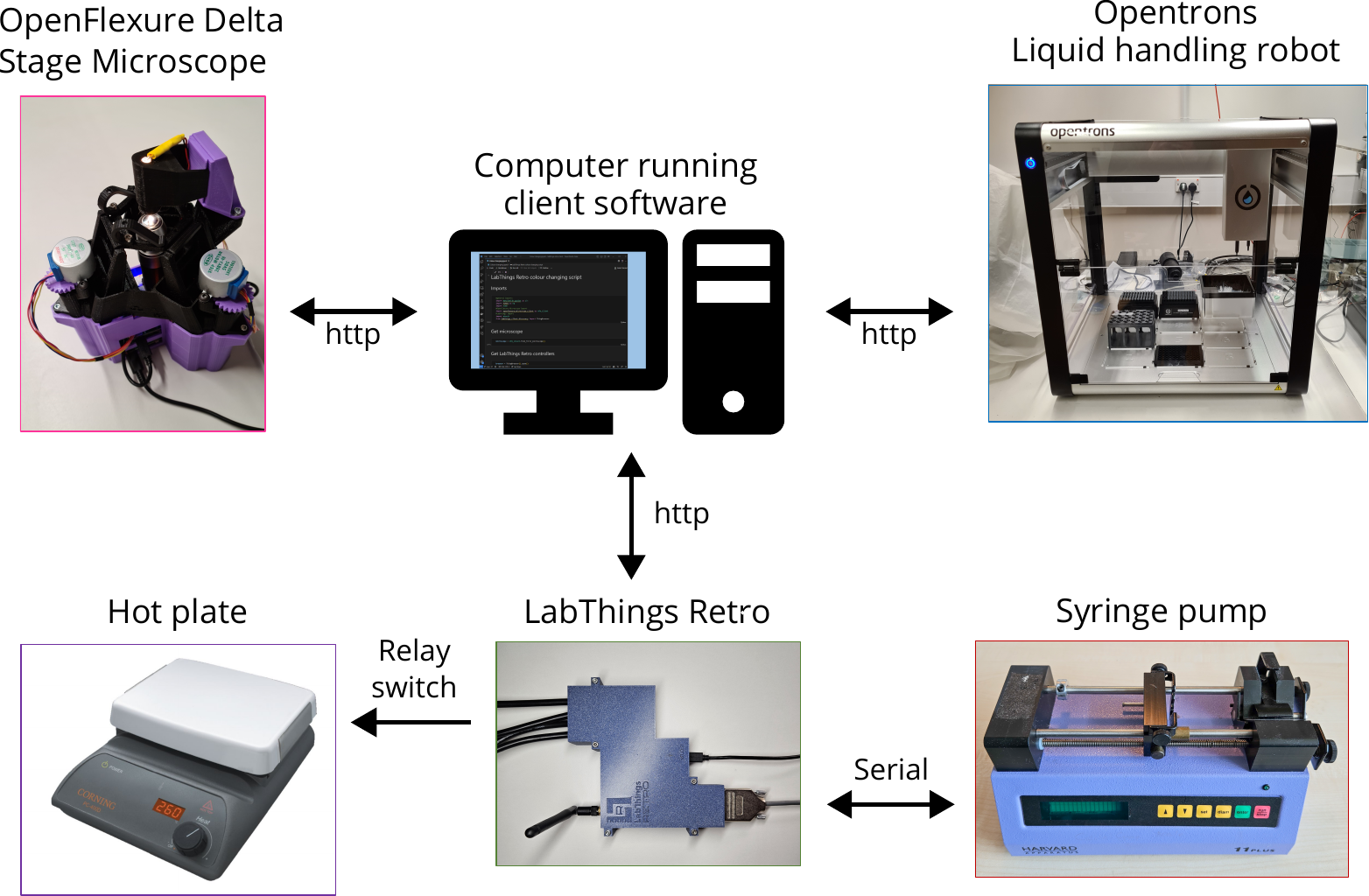}
    \caption{An example smart lab layout for automated experiments. A computer running client software is able to communicate with a range of scientific instruments with modern communications standards (top). In addition to issuing commands, it can receive data from the instruments and make decisions about how to progress the experiment.  LabThings Retro enables older, but still functioning, equipment (bottom) to be integrated into this modern experimental setup.}
    \label{fig:smart-lab}
\end{figure}

In this paper, we describe \textit{LabThings Retro}, our solution to integrating and retrofitting devices in categories 1 and 2 above, to allow automated smart experiments.  An example of a smart lab using this design is shown in Figure~\ref{fig:smart-lab}:  one computer (which could be in a separate building to the experiments), using a client written in the language of the user's choosing, is able to communicate to both new (top) and old  devices (bottom) using the established web-of-things standard. Communication  with older devices is made possible using the LabThings Retro controller.  

This work demonstrates how it is possible to integrate  older devices into modern smart lab experiments, enabling laboratories globally to reuse equipment, save money, prevent e-waste, standardize protocols, and empower researchers to tackle previously daunting experimental challenges, potentially unlocking knowledge that is only accessible via automated or high throughput approaches.

\section{Design requirements}
The purpose of the LabThings Retro controller is to enable the  use of older laboratory devices with new web communication standards. Our objectives are the following:
\begin{itemize}
    \item Develop a simple controller which can be easily replicated. It should be possible to manufacture using 3D printers and to be assembled using hand tools, following an easy-to-use manual.
    \item The controller, including electronics, should be available to manufacture at a low cost and be open source. 
    \item The controller should be able to connect to a network for remote control.
    \item The controller should be able to control devices which can be turned on and off with a switch e.g. light sources, stirrers or heaters.
    \item The controller should be able to send and receive more complex commands and data over a serial connection to e.g. syringe pumps, incubators, or pH meters.
\end{itemize}      

\section{Hardware Design}
The hardware was designed to be easy to replicate and modify. The example we provide works with any Category 1 devices as defined in the introduction, and can be adapted to work with any Category 2 device that is controlled using serial communication protocols.  

\begin{figure}
    \centering
    \includegraphics[width=\linewidth]{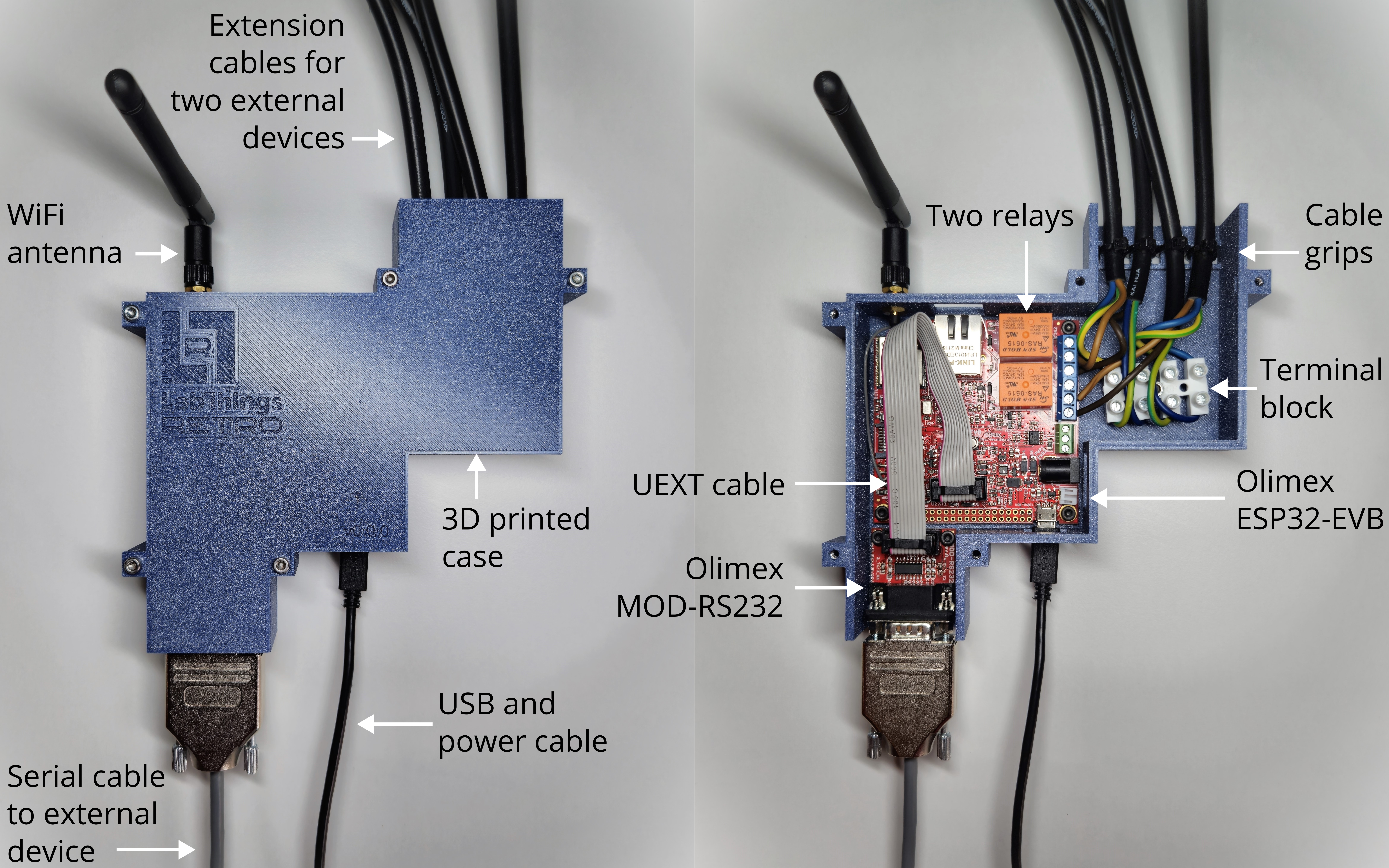}
    \caption{LabThings Retro hardware design. Left: The device with the lid on, showing the external connections.  Right: The device with the lid off, showing the internal components.}
    \label{fig:labthings-retro-hardware}
\end{figure}

\subsection{Electronics}

\subsubsection{Microcontroller board}
For the microcontroller board, we evaluated ESP32-based boards.  This is because ESP32 is a well-documented and versatile microcontroller with a good developer ecosystem around it.  It also has WiFi built-in and sufficient CPU and memory. In addition, a LabThings library already exists for this architecture~\cite{noauthor_github_nodate-2}, based on the WebThingsIO webthing-arduino library~\cite{noauthor_github_nodate}.

There are several boards built for the ESP32. We wanted a board which had relays to turn laboratory equipment on and off. In addition, the board should be able to communicate with common serial protocols, such as RS232. Finally, we looked for a board which was open source, to enable better documentation, community support and longevity of the device. Therefore we decided on the Olimex ESP32-EVB~\cite{olimex_esp32-evb_nodate}. This ESP32-based board has a number of connection options including Wifi, Bluetooth Low Energy (BLE), and an ethernet port.  It has two 10A/250VAC relays to switch externally powered devices, can connect to Olimex's range of UEXT modules (such as serial interfaces) via a UEXT connector and is certified as OSHWA Open Source Hardware with UID BG000011~\cite{noauthor_oshwa_nodate}.   

\subsubsection{Serial communication interface}
To control external serial devices, we used the Olimex MOD-RS232~\cite{olimex_mod-rs232_nodate}.  This module connects to the ESP32-EVB microcontroller board with UEXT. It has an RS232 level-shifter to enable serial communications from the ESP32 to serial devices with a DB-9 connector operating at the standard 5\,Volt level. It is also possible to use the Olimex MOD-RS485 for communication with RS485 and RS422 serial devices~\cite{olimex_mod-rs485_nodate}.

\subsubsection{Electronic relays}
Many devices in the lab are Category 1 devices (as defined in the introduction) and just need to be turned on or off.  For example: lamps, heaters, or shakers. The Olimex ESP32-EVB has two 10A/250VAC relays, which can power on or off two connected devices.     

\subsection{3D printing}
To contain the electronics and make the device safe with the relay circuit, the device is contained within a 3D-printed enclosure. The two-part case can be printed with simple filament deposition 3D printers using standard materials. If using the relay circuits, we recommend using a UL94V-0 flame retardant ABS. The enclosure is designed using OpenSCAD, and so can be easily adapted to other configurations. The files are available on the GitLab repository~\cite{noauthor_labthingscase_2023}.

\subsection{Assembly}
An important aspect of the design is the ease of assembly.  A thorough set of assembly instructions, powered by gitbuilding~\cite{stirling_gitbuilding_2022}, contains photos and a Bill of Materials. The device can be assembled using simple hand tools and common hardware. The assembly instructions are available on the GitLab website~\cite{mcdermott_labthingshome_nodate}.

\begin{figure}[t!]
    \centering
    \includegraphics[width=\linewidth]{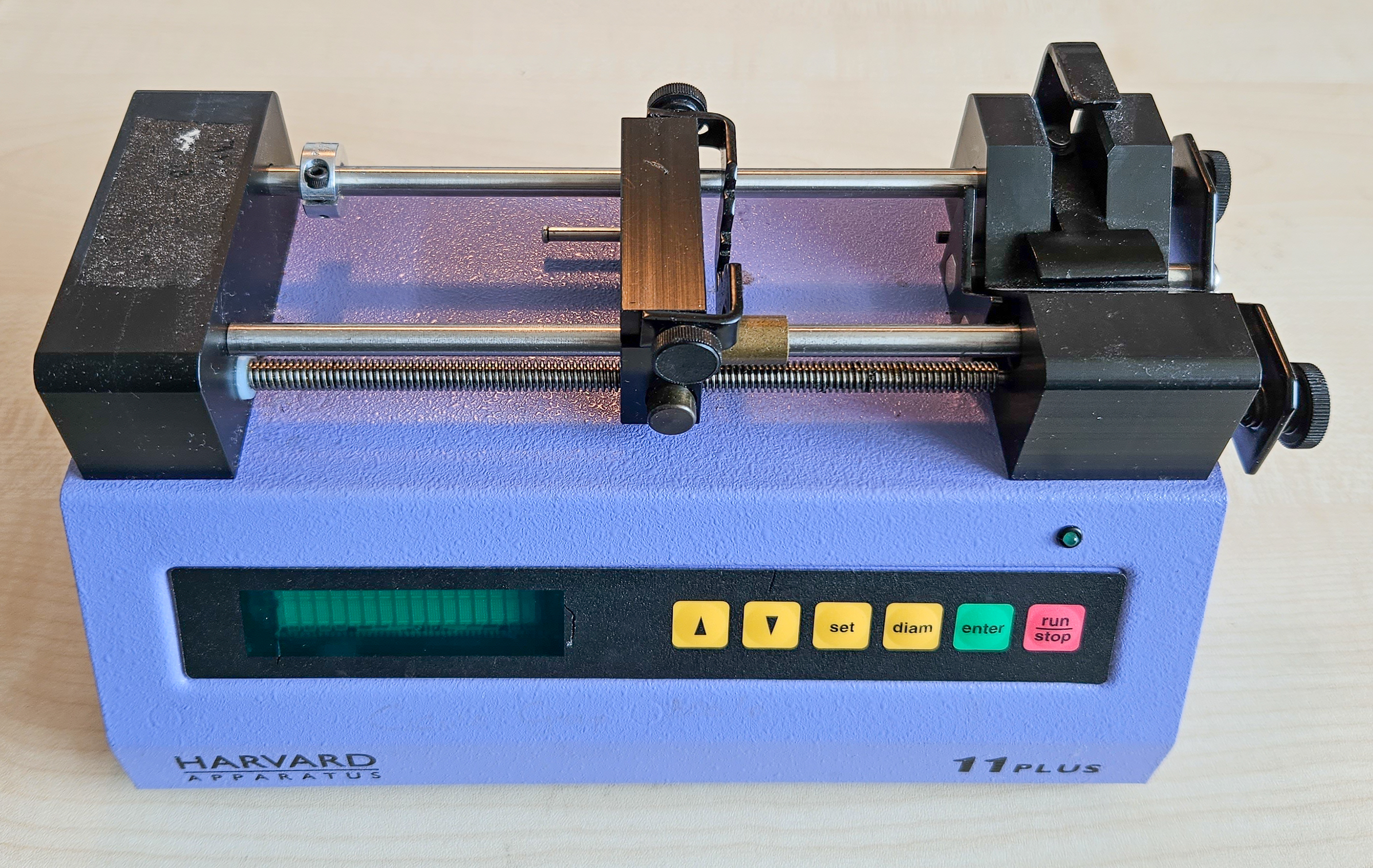}
    \caption{The Harvard Apparatus \textit{11 Plus syringe pump} is used as an exemplar Category 2 serial device.  As well as the on-device buttons and screen, it can be controlled using the RS-232 serial protocol.}
    \label{fig:syringe-pump}
\end{figure}

\section{Software Design}
We have developed example software for using the device to control the relays and to send commands over the serial interface.  For the exemplar serial device, we chose a Harvard Apparatus \textit{11 Plus syringe pump} (see Figure~\ref{fig:syringe-pump}). Syringe pumps are used for many biological experiments, for example, to perfuse liquids in cell culture, or to  control flow through microfluidics. They are long-lasting and typically  reliable pieces of laboratory equipment, so older syringe pumps typically remain functional for use in experiments.  However, as they use older serial communication protocols, it is difficult to control them with modern computers. For example, most modern computers do not have the required serial ports, drivers, and software for the devices are old and may not work with current operating systems.

\subsection{LabThings}
For the controller software, we built upon the LabThings ESP32 library~\cite{noauthor_github_nodate-2}. LabThings is a project that applies the W3C Web of Things (WoT) standard~\cite{noauthor_home_nodate} to laboratory instruments, providing Python client and server libraries and a number of examples~\cite{noauthor_labthings_nodate}.

It was originally created for the OpenFlexure Microscope~\cite{collins_simplifying_2021} but lends itself well to generalisation. LabThings aims not to define new standards, but to make use of existing, well-supported technologies already in widespread use for Internet of Things technologies. In particular, HTTP Application Programming Interfaces (APIs) allow applications to communicate over a network. The OpenAPI standard provides human and machine-readable documentation of this API, allowing easy control and even code generation in most programming languages. The Thing Description (part of the W3C WoT standard) describes a device's capabilities at a higher level. There are many well-tested libraries and tool kits to support these technologies due to their widespread use outside of scientific laboratories.

\subsection{Serial communication}
When two devices communicate using serial communication protocols, each data bit is sent sequentially over a single data line. In contrast, in parallel communication, multiple bits are sent simultaneously over several wires. Serial links are more commonly used for device communications, due to their simplicity and lower cost of implementation. Modern serial interfaces such as USB are well known due to their use in everyday electronic products. Older serial interfaces such as RS-232 and RS-485 are commonplace in industrial and scientific instruments due to their wide adoption and ease of implementation. However, such interfaces are rarely present on modern computers, and low-level adapters providing serial ports over USB suffer from difficulties around device discovery and consistent addressing.

Our exemplar device, the syringe pump, can be controlled remotely using serial communication.  It uses the RS-232 interface and the manual describes the commands, queries and responses. These are sent as ASCII codes over the RS-232 connection. Each command consists of a string of ASCII characters. The first two characters are the address of the pump. This is included for when you are using multiple pumps in a daisy chain, and the address of the pump is set on the pump itself.  In our example, it is \verb|00|. The next characters identify the command.  For example, to start running the pump we send \verb|RUN|. The message is terminated with a carriage return character \verb|\r|. The full char array \verb|00RUN\r| is then sent to the pump. A prompt is returned from the pump, indicating its status, such as running, or stopped.  It is also possible to send commands with numbers, for example, the flow rate, or the target infusion volume. To get information back from the pump, such as the current infused volume, it is possible to send queries.

Upgrading a serial device with LabThings to the WoT standard can be achieved in two ways: 
\begin{enumerate}
    \item \textbf{A network-to-serial converter}, in which serial commands are sent to the LabThings Retro controller over HTTP, and are then relayed to the device's communications port. Replies would then be returned to the client in the HTTP response. Operating in this mode has the advantage of immediately exposing all the device's capabilities to the client, and using one firmware image to control a variety of different devices. However, it would push more complexity into the client code: the device's capabilities are not exposed in the HTTP API or the Thing Description, so the user would still need to look up command codes in the device manual. The low-level interface also means the device could behave oddly if multiple clients connect, and so many of the limitations of a serial connection still apply.
    \item \textbf{A firmware image that exposes the device's commands as HTTP endpoints}, documented as WoT Actions and Properties. This would require more effort to develop, but is much easier to use: all capabilities of the device are documented in the API, so client code can be generated automatically and the user need not refer to the device's manual. Using a higher-level interface over the network also means that the LabThings Retro controller can ensure communications do not become confused - for example ensuring that each command receives the correct response, and preventing commands from being sent when the instrument is not ready to receive them. This confers a level of robustness to concurrent use by multiple clients, which can be helpful when monitoring an ongoing experiment.
\end{enumerate}

\begin{figure}
    \centering
    \includegraphics[width=\linewidth]{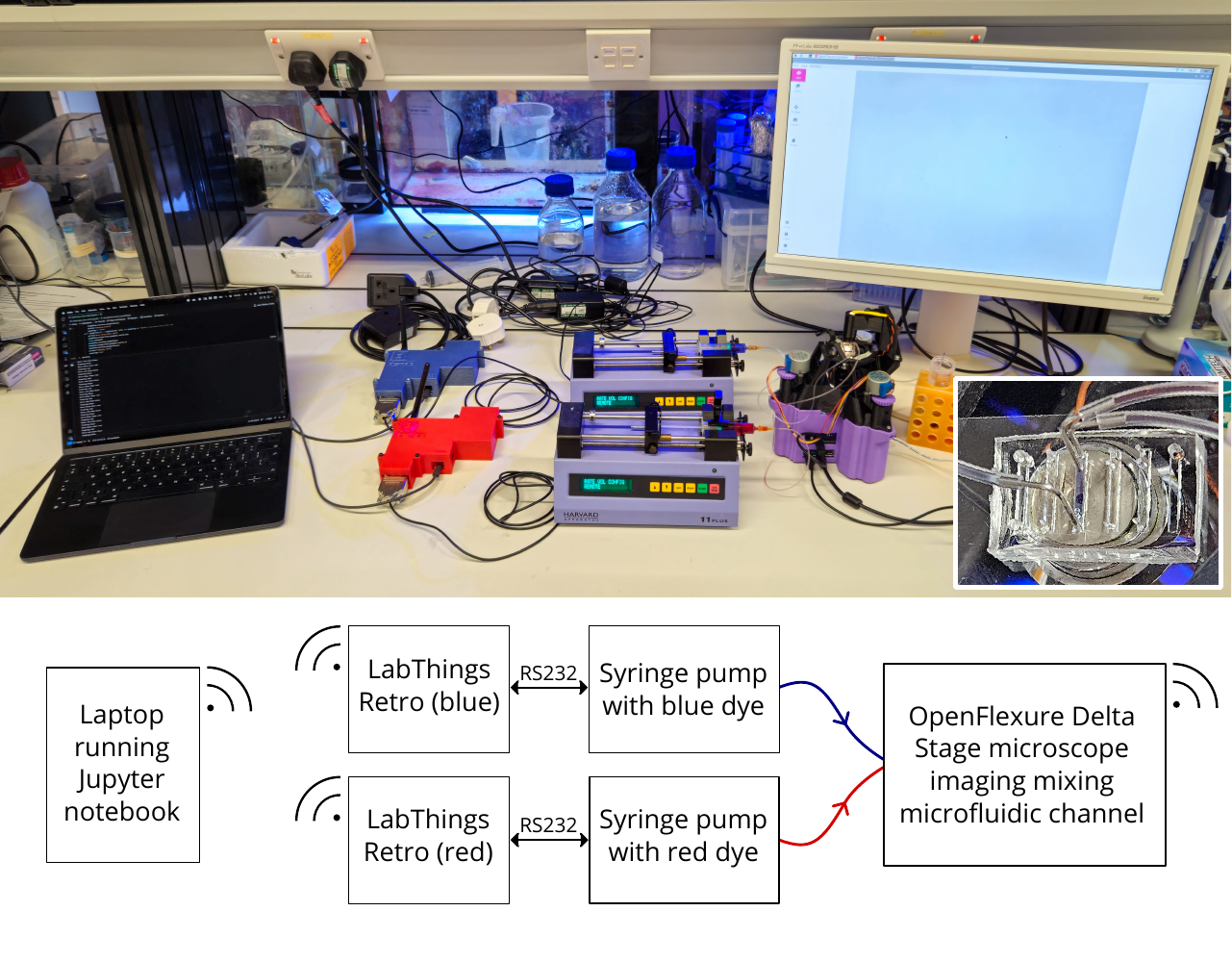}
    \caption{Experimental setup for demonstrating a closed feedback control across various devices using LabThings Retro. A laptop running a Jupyter notebook is connected to two LabThings Retro controllers over WiFi.  The laptop is also connected to an OpenFlexure Delta Stage microscope over WiFi.  Each LabThings Retro controller is connected to a syringe pump, one with a syringe containing blue dye and the other holding a syringe with red dye. The dyes pass along two tubes and are mixed in a microfluidic channel (insert). The microfluidic channel is imaged using the OpenFlexure Delta Stage microscope.}
    \label{fig:experimental-setup}
\end{figure}

\begin{figure}
    \centering
\begin{tikzpicture}[node distance = 1.5cm]

\node (start) [startstop] {Start};
\node (initParams) [process, below of =start] {Initialise syringe pumps' parameters (e.g. syringe volume, inner diameter)};
\node (infusePump) [process, below of = initParams] {Infuse first pump at high flow rate.};
\node (getFrame) [io, below =0.5cm of  infusePump]{Get frame from microscope.};
\node (colourChange)[decision, below = 0.5cm of getFrame] {Has current frame changed colour?};
\node (reduceRate)[process, below = 1cm of colourChange] {Reduce current pump's flow rate for 5 seconds.};
\node (stopCurrent)[process, below of =reduceRate]{Stop current pump.};
\node (startNext)[process, below of = stopCurrent]{Infuse other pump at high flow rate.};

\draw [arrow] (start) -- (initParams);
\draw [arrow] (initParams) -- (infusePump);
\draw [arrow] (infusePump) -- (getFrame);
\draw [arrow] (getFrame) -- (colourChange);
\draw [arrow] (colourChange) -- node[anchor=north]{No} ++(-3,0) |- (getFrame);
\draw [arrow] (colourChange) -- node[anchor = east]{Yes} (reduceRate);
\draw [arrow] (reduceRate) -- (stopCurrent);
\draw [arrow] (stopCurrent) -- (startNext);
\draw [arrow] (startNext) -- ++(3,0) |- (getFrame);

\end{tikzpicture}
\caption{The algorithm flow for the automation case study. The result of this algorithm is that the microscope's field of view oscillates between red and blue. Once the algorithm detects that one pump has changed the colour in the microscope's field of view, it reduces the pump's flow rate for 5\,seconds before turning off that pump and starting the other.}
\label{fig:algorithmflow}
\end{figure}
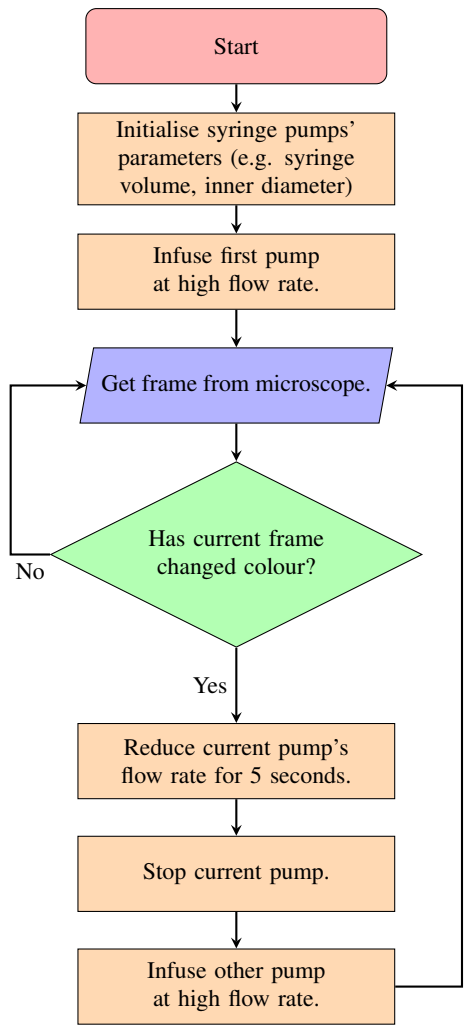

Our exemplar syringe pump implementation uses the second of these approaches. As such, the LabThings Retro controller documents all the capabilities of the syringe pump in the API. When the LabThings Retro controller receives a WoT Action or Property request, it converts that request into the relevant serial command and sends it to the device. It then returns the result to the client.

\subsection{Electronic relays}
The state of each electronic relay is a Boolean property. When the server receives a WoT Property set request from the client, it updates the relay switch accordingly.  The client can also get the current value of the property to determine whether the relay is on or off. 

\subsection{Installation}
The full instructions for installation can be found in the assembly instructions~\cite{mcdermott_labthingshome_nodate}. To install and configure the software, we use PlatformIO~\cite{platformio_platformio_nodate}. The example server code can be downloaded from the GitLab repository~\cite{noauthor_labthingsserver_2023}. Once downloaded, the LabThings library can also be installed.  The user will need to add their WiFi hot-spot SSID and password into the code so that it can be controlled from other devices on the same network. For security reasons, it is recommended that this WiFi hotspot is not connected to the outside internet. The code can then be flashed to the Olimex ESP32-EVB and the device is ready to use.

\subsection{Client}
The advantage of controlling LabThings Retro controllers using the LabThings library is that it extends the WebThings Library. As this open standard is self-documenting, it is not necessary to code clients to work with individual devices.  There are therefore a range of client libraries that have been designed to work on a variety of devices. It is also possible to create a library for any language that has HTTP support or to automatically generate one based on the OpenAPI description provided by the LabThings Retro server. This makes it easy for users to build scripts in their preferred language on the internet-enabled device of their choice. For example, LabThings Retro controllers can be included within existing MATLAB or Python scripts.

Scripts can be easily developed to control several devices simultaneously. Scripts are useful for automating complex experiments, such as parameter sweeps, where manual control of equipment would be overly time-consuming, and they are essential in more complex automation that aims to deploy feedback and respond to events in the experiment itself. Example client scripts and installation instructions are available in the GitLab repository~\cite{noauthor_labthingsclient_2023}.

It is also possible to control LabThings Retro controllers using Graphical User Interfaces (GUIs). WebThings Gateway~\cite{noauthor_github_nodate-1} creates a web interface that can automatically control LabThings Retro controllers, as well as other open and commercial communication protocols through add-ons.

\section{Automation case study}
To demonstrate the potential use of LabThings Retro, we designed the experiment shown in 
Figure~\ref{fig:experimental-setup}. This experiment combines two LabThings Retro controllers controlling two syringe pumps and an OpenFlexure Delta Stage microscope~\cite{McDermott2022}. All three devices are controlled remotely using a Python script~\cite{noauthor_labthingsclient_2023}. This experiment demonstrates a smart experiment, made possible using LabThings Retro, which creates an autonomous feedback loop between the new microscope and the older syringe pumps.

Two coloured dyes (one red and one blue) were diluted 1:10 in water.  They were drawn into two 5\,ml syringes and inserted into the two syringe pumps. The two syringes were connected to a PDMS microfluidic mixing channel, which was positioned on the microscope. 

Each LabThings Retro controller was connected to its corresponding syringe pump with an RS232 serial cable.  The two LabThings Retro controllers were connected to the same WiFi network as the laptop and OpenFlexure Delta Stage. A simple Python script was written to run the automation. The algorithm flow is shown in Figure \ref{fig:algorithmflow} and a video of the experiment is in the supplementary material.

This completely autonomous algorithm takes input from the microscope, makes a decision based on the contents of the frame and controls the syringe pumps accordingly. LabThings Retro changes the operational mode and settings of the syringe pumps during the experiment, for example reducing the flow rate. It could therefore easily be scaled up to control more equipment or use advanced computer vision techniques in order to develop a smart biological experiment.

\section{Conclusion}
In this paper, we have demonstrated how it is possible to integrate older equipment into modern smart experiments with LabThings Retro. This will enable more laboratories to develop automated experiments. Our solution is low-cost, easily modifiable and can be extended to a range of devices and experiment types.

LabThings Retro has the advantage that it is built on an established protocol for communicating with devices. There already is a range of clients and GUIs available for controlling devices based on WoT protocols. This means that it is not necessary for the control scripts to use the same framework that is used in the instrument code. 

In future work, we will extend the networking capabilities of the device by developing a LabThings ethernet adapter to use the ESP32-EVB's ethernet port. To assist with the adoption of this smart lab approach, the sharing of code for controlling particular instruments with the LabThings Retro framework would make it quicker to create experiments.  

\bibliographystyle{IEEEtran}
\bibliography{LabThingsRetro.bib}

\end{document}